\documentstyle[prd,aps,preprint,tighten,eqsecnum,epsf,amssymb]{revtex} 
\newcommand{\naught}{{\scriptscriptstyle 0}}

\newcommand{\thop}{\eth}
\newcommand{\thopnaught}{\thop{}_{\naught}}
\newcommand{\thopbarnaught}{\bar{\thop}{}_{\naught}}
\newcommand{\referencek}{{\mathsf k}}
\newcommand{\referencel}{{\mathsf l}}
\newcommand{\referenceu}{{\mathsf u}}

\newcommand{\referenceK}{{\mathsf K}}
\newcommand{\referencerho}{{\varrho}}
\newcommand{\referencesigma}{{\varsigma}}
\newcommand{\referenceSigma}{{\mathsf \Sigma}}

\newcommand{\referenceN}{{\mathsf N}}
\newcommand{\referenceM}{{\mathsf M}}
\newcommand{\referenceA}{{\mathsf A}}

\begin{document}
\preprint{{\tt gr-qc/9810003}}

\title{Canonical Quasilocal Energy and Small Spheres}

\author{J.~D.~Brown,\footnote{Department of Physics, North Carolina
State University, Raleigh, NC 27695--8202 USA}\footnote{Department
of Mathematics, North Carolina State University, Raleigh,
NC 27695--8205 USA} 
S.~R.~Lau,\footnote{Department
of Physics \& Astronomy, University of North Carolina,
CB\# 3255 Phillips Hall, Chapel Hill, NC 27599-3255 
USA}\footnote{Institut f\"{u}r Theoretische
Physik, Technische Universit\"{a}t Wien, Wiedner   
Hauptstra{\ss}e 8-10,
A-1040 Wien, \"{O}sterreich}\footnote{Current address:
Applied Mathematics Group,
Department of Mathematics,
University of North Carolina, CB\# 3250 Phillips Hall,
Chapel Hill, NC 27599-3250 USA} and
${\rm J.~W.~York}^{*\ddag}$}

\maketitle
\begin{abstract}
Consider the definition $E$ of quasilocal energy stemming
from the Hamilton-Jacobi method as applied to the canonical
form of the gravitational action. We examine $E$ in the
standard ``small-sphere limit,'' first considered
by Horowitz and Schmidt in their examination of
Hawking's quasilocal mass. By
the term {\em small sphere} we mean a cut $S(r)$, level in an
affine radius $r$, of the lightcone $N_{p}$ belonging to a
generic spacetime point $p$. As a power series in $r$, we
compute the energy $E$ of the gravitational and matter fields
on a spacelike hypersurface $\Sigma$ spanning $S(r)$.
Much of our analysis concerns conceptual and technical issues
associated with assigning the zero-point of the energy. For
the small-sphere limit, we argue that the correct zero-point
is obtained via a ``lightcone reference,'' which stems from
a certain isometric embedding of $S(r)$ into a genuine
lightcone of Minkowski spacetime. Choosing this
zero-point, we find the following results: (i)
in the presence of matter
         $E = {\textstyle \frac{4}{3}} \pi r^{3}
              [ T_{\mu\nu} u^{\mu} u^{\nu}]|_{p}
            + {\rm O}(r^{4})$ and (ii)
in vacuo
         $E = {\textstyle \frac{1}{90}} r^{5}
              [T_{\mu\nu\lambda\kappa}
              u^{\mu} u^{\nu} u^{\lambda} u^{\kappa}]|_{p}
            + {\rm O}(r^{6})$.
Here, $u^{\mu}$ is a unit, future-pointing, timelike vector
in the tangent space at $p$ (which defines the choice of affine
radius); $T_{\mu\nu}$ is the matter stress-energy-momentum
tensor; $T_{\mu\nu\lambda\kappa}$ is the Bel-Robinson
gravitational super stress-energy-momentum tensor; and
$|_{p}$ denotes ``restriction to $p$.''
Hawking's quasilocal mass expression agrees with the results
(i) and (ii)  up to and including the first
non-trivial order in the affine radius. The non-vacuum result
(i) has the expected form based on the results of Newtonian
potential theory.
\end{abstract}

\newpage

\section*{Introduction}
Consider Einstein's non-covariant first-order action\cite{York},
the 4-integral of a ``bulk'' Lagrangian which is quadratic in
the Christoffel symbols and thus often called the
``$\Gamma\Gamma$ action.'' Starting with the Einstein action,
one applies standard techniques associated with
Noether's theorem in
order to derive, among other things, an energy definition in
general relativity: namely, the 2-integral of an
Einstein superpotential over some generic
2-surface $S$ in spacetime $M$.\footnote{See the excellent
review article by Goldberg in Ref.~\cite{Goldberg1} for the
details of this analysis.} The Einstein energy is well
known to be ambiguously defined because it depends
on the choice of background coordinates. Nevertheless, one may
use the Einstein construction to define sensible notions of
total gravitational energy. Indeed, consider the scenario of
asymptotic flatness, say, towards
future null infinity $J^{+}$. In this case, $S$ tends to a
round, infinite-radius, 2-sphere cut of $J^{+}$, and the
(now suitably unique) choice of asymptotically Cartesian
coordinates ensures that the Einstein energy agrees with the
accepted Trautman-Bondi-Sachs ({\sc tbs}) notion of total
energy.\cite{Goldberg1}  However, were we to offer the
Einstein definition as the energy contained within some
{\em quasilocal} (that is, finite) 2-surface $S$, we would
still be confronted with the
task of choosing a physically meaningful
set of background coordinates. The only natural choice would
be coordinates which are partially adapted to the
embedding of $S \subset M$. However, such a choice wrecks the
agreement between the Einstein and {\sc tbs} energies as $S$
tends towards $J^{+}$. In fact, choosing such $S$-adapted
coordinates, one finds that the Einstein
energy blows up in the said limit.
Similar statements can be made regarding other approaches for
defining energy, which trade coordinate (that is,
holonomic-frame) ambiguity for ambiguity of a different stripe,
e.~g.~tetrad (or rigid) frame, spin frame, or auxiliary vector
(or spinor) fields. The traditional party line regarding these
issues is the following: {\em there is no over-arching rule,
applicable for all quasilocal 2-surfaces, for selecting a
(suitably unique) background frame}; whence gravitational
quasilocal energy is not well-defined.\footnote{A serious
contender for such an over-arching rule has been given by
Dougan and Mason, who use certain ``(anti-)holomorphic''
spinor fields in order to define a four-dimensional space of
``quasi-translations'' pointing on and off an essentially
generic 2-surface.\cite{DouganMason} Szabados
has shown that the Dougan-Mason proposal provides a tidy
characterization of pp-wave geometries.\cite{Szabados}}
To what extent does the stubborn presence of frame ambiguity
in the quasilocal context point to a gap in our understanding
of gravitational energy? To address this question and sharpen
our thoughts on these issues, let us consider a covariant 
version of the Einstein construction.

Employing a straightforward field-theoretic generalization of
the Hamilton-Jacobi ({\sc hj}) method\cite{Lanzcos}, one may
derive from a covariant action functional an expression
for canonical quasilocal energy  ({\sc qle})
in general relativity.\cite{BrownYork} We call this definition of {\sc
qle}
{\em canonical}, because, owing to its intimate connection
with {\sc hj} theory, this {\sc qle} is also the on-shell value
of the gravitational Hamiltonian for the choice of
unit lapse function and vanishing shift vector
at the system boundary.\cite{BrownYork}
We also note that the canonical {\sc qle} is the
thermodynamic internal energy in a thermodynamical description
of a (relativistic) self-gravitating system.\cite{thermopapers}
The analysis that leads to the canonical {\sc qle}
runs along a somewhat
different line than the one followed in a Noether-type
analysis, but it also leads to a
concept of energy which is not unique. Indeed,
{\em as is always the case with energy}, the
canonical {\sc qle} is defined only
up to the choice of a zero-point. The
zero-point ambiguity may be traced to a freedom present in the
action principle. Namely, one may always add to the action
any functional of the fixed boundary data without affecting the
variational principle. As with the situation above, if the
goal is to obtain agreement with the accepted notions of
gravitational energy at spacelike or null infinity, then there is
a suitably unique choice of
energy zero-point\cite{BrownYork,BrownLauYork},
whereas at the quasilocal level there seems
to be no preferred choice.\footnote{Other than the choice of a
vanishing zero-point. This choice corresponds to the
aforementioned choice of $S$-adapted coordinates for the
Einstein definition. Like before, such a choice, leading to an
infinite energy, wrecks the agreement between the
canonical quasilocal
and {\sc tbs} energies in the large-sphere null limit.}
While at first sight this seems no better or worse than the
situation encountered above, notice that now the ambiguity in
the energy has a physical interpretation, and, moreover, is a
field-theoretic generalization of the
standard ambiguity present
in the {\sc hj} definition of energy in ordinary mechanics.
We may now restate the emphasized portion of the party line above
as follows: {\em there is no over-arching rule, applicable for
all quasilocal 2-surfaces, for selecting a (suitably unique)
zero-point.} Taking this statement at face value, we claim
that it is the physicist's job to select an appropriate
energy zero-point, guided by the principle that the
selection should be appropriate for the physics of the
problem at hand. We would
like to point out that this is a common enough state of
affairs in general relativity, a many-faceted theory known
for its wealth of possible boundary conditions. Indeed, by
way of analogy consider the search for solutions of
the Einstein field equations. In practice, relativists
certainly do not attempt to find {\em the} general
solution, rather they attempt to find solutions given
some additional physical input (boundary conditions,
symmetries, etc.). In practice, the same such additional input
is needed to associate a meaningful {\sc qle} with a
particular quasilocal 2-surface.

Bearing these points in mind, we
recall the form of the canonical {\sc qle}:
\begin{equation}
E = (8\pi)^{-1} \int_{S} {\rm d}S
                \left(k -
                      k|^{\rm {\scriptscriptstyle ref}}
                \right)
{\,},
\label{BrownYorkqle}
\end{equation}
where we adopt geometrical units (in which both Newton's
constant and the speed of light are set to unity),
$S$ is a closed 2-surface, ${\rm d}S$ is the proper area
element on $S$, and $k$ is the mean curvature of $S$ as
embedded in a spanning 3-surface $\Sigma$. It
is important to realize that this $E$, while obtained as a
proper surface integral over $S$, is the energy of the
gravitational and matter fields which live on $\Sigma$, that is
to say, $E$ is a functional of the initial data of $\Sigma$. This
concept of energy is rooted in the $3+1$ view of spacetime
geometry, and for a fixed $S$ it is slightly sensitive to the
choice of spanning $\Sigma$. [More precisely, $E$ depends on
the equivalence class of spanning 3-surfaces determined by a
unit timelike vector on (and pointing orthogonal to) $S$.]
This sensitivity is quite analogous to the observer dependence
of energy in special relativity, and {\em a priori} we expect
its presence.\cite{BrownLauYork2} In general, the term
$k|^{\rm {\scriptscriptstyle ref}} =
k|^{\rm {\scriptscriptstyle ref}}(\sigma_{ab})$ represents an
arbitrary (local) function of the intrinsic metric
$\sigma_{ab}$ of $S$ and
corresponds to the freedom to assign the {\sc qle}
zero-point. Notice that this freedom corresponds to a proper
surface integral of what is effectively a free function of two
variables. This freedom, stemming from the field-theoretic
character of gravity, is rather more subtle than the
freedom in simple mechanics of simply adding a constant to
the energy. For our analysis here we only consider
two energy zero-points,
one determined by {\em lightcone reference} and the other
by {\em Euclidean reference}.
We make these concepts precise below.

In this paper we examine the canonical {\sc qle} in the
standard ``small-sphere limit,'' first considered by Horowitz
and Schmidt in their classic examination of Hawking's
quasilocal mass.\footnote{This limit is also considered in
Ref.~\cite{KTW} by Kelly, Tod, and
Woodhouse for Penrose's kinematic twistor and associated
quasilocal mass, and in Ref.~\cite{Dougan} by
Dougan for the Dougan-Mason quasilocal four-momentum.} By the
term {\em small sphere} we mean a cut $S(r)$, level in an
affine radius $r$, of the lightcone $N_{p}$ belonging to a
generic spacetime point $p$. As a power series in $r$, we
compute the energy $E$ of the gravitational and matter fields
on a spacelike hypersurface $\Sigma$ spanning $S(r)$. Much
of our analysis concerns conceptual and technical issues
associated with assigning the zero-point of the energy, and,
therefore, particularly elucidates the points raised in the
first two paragraphs above. For
the small-sphere limit, we argue that the correct zero-point
is obtained via the aforementioned
lightcone reference, which stems from
a certain isometric embedding of $S(r)$ into a genuine
lightcone of Minkowski spacetime and amounts to fixing
\begin{equation}
       k |^{\scriptscriptstyle {\rm ref}}
=    - \sqrt{ {\textstyle \frac{1}{2}}
       {\cal R}}
       \left\{1 + {\textstyle \frac{1}{6}}
       ({\cal R})^{-1}\Delta\log
       {\cal R} + \left[1 + {\textstyle \frac{1}{6}}
       ({\cal R})^{-1}\Delta\log
       {\cal R}\right]\!{}^{-1}\right\}
\label{lcmeancurvature}
\end{equation}
in the above {\sc qle} definition. Here $\Delta$ is the Laplacian
operator and ${\cal R}$ is twice the Gaussian curvature of $S(r)$;
therefore, as advertised,
this choice for $k |^{\scriptscriptstyle {\rm ref}}$
depends solely on the intrinsic geometry of
$S(r)$. Notice that,
due to the presence of the square root in this expression, one
expects this choice for $k |^{\scriptscriptstyle {\rm ref}}$ to
be valid only for (topologically spherical) 2-surfaces
possessing everywhere positive Gaussian
curvature (as is the case both for the small spheres we study
here). Choosing the proper surface integral of this choice for
$(8\pi)^{-1} k |^{\scriptscriptstyle {\rm ref}}$ as the energy
zero-point, we find the following results for small spheres: in
the presence of matter
\begin{equation}
          E = {\textstyle \frac{4}{3}} \pi r^{3}
              [ T_{\mu\nu} u^{\mu} u^{\nu}]|_{p}
            + {\rm O}(r^{4})
\label{result_one}
\end{equation}
and in vacuo
\begin{equation}
          E = {\textstyle \frac{1}{90}} r^{5}
              [T_{\mu\nu\lambda\kappa}
              u^{\mu} u^{\nu} u^{\lambda} u^{\kappa}]|_{p}
            + {\rm O}(r^{6}) {\,}
\label{result_two}
\end{equation}
Here, $u^{\mu}$ is a unit, future-pointing, timelike vector
in the tangent space at $p$ (which defines the choice of affine
radius); $T_{\mu\nu}$ is the matter stress-energy-momentum
tensor; $T_{\mu\nu\lambda\kappa}$ is the Bel-Robinson
gravitational super
stress-energy-momentum tensor;\footnote{Our vacuum-case
                             Bel-Robinson tensor is the
                             following:
                             \begin{equation}
                             T_{\mu\nu\lambda\kappa} :=
                             C_{\mu\rho\lambda\sigma}
                             C_{\nu}{}^{\rho}{}_{\kappa}
                             {}^{\sigma} +
                             {}^{*}\:\!\!
                             C_{\mu\rho\lambda\sigma}
                             {}^{*}\:\!\!C_{\nu}
                             {}^{\rho}{}_{\kappa}{}^{\sigma}
                             = 4 {\,} {}^{\scriptscriptstyle +}
                             \:\!\! C_{\mu\rho\lambda\sigma}
                             {}^{\scriptscriptstyle -}\:\!\!
                             C_{\nu}{}^{\rho}{}_{\kappa}
                             {}^{\sigma}
                             {\,} ,
                             \label{BelRobinson}
                             \end{equation}
                             where $C_{\mu\rho\lambda\sigma}$
                             is the
                             Weyl tensor,
                             ${}^{*}\:\!\!
                             C_{\mu\rho\lambda\sigma}
                             := \frac{1}{2}
                             \epsilon_{\mu\rho\alpha\beta}
                             C^{\alpha\beta}{}_{\lambda\sigma}$
                             is the left-dual of the Weyl
                             tensor, and
                             ${}^{\scriptscriptstyle \pm}
                             C_{\mu\rho\lambda\sigma}
                             := \frac{1}{2}
                             (C_{\mu\rho\lambda\sigma} \mp
                             \frac{1}{2}{\rm i}
                             \epsilon_{\mu\rho\alpha\beta}
                             C^{\alpha\beta}{}_{\lambda\sigma})$
                             is the self-dual $(+)$
                             [anti-self-dual $(-)$]
                             part of the Weyl tensor.
                             Further curvature
                             conventions are discussed in the
                             appendix.}
and $|_{p}$ denotes ``restriction to $p$.''
It is interesting to note that, when integrated, the 
Bel-Robinson ``energy'' in
(\ref{result_two}) has been proven to be very useful already
in the sense of a mathematical ``energy'' in the study of existence
of solutions of hyperbolic equations in 
Einsteinian gravity.\cite{Friedrich,CK,ACY,York2} It is noteworthy
that it shows up also in the physical limit given in (\ref{result_two}).
Although the full physical significance of the physical limit is
not known to us, we think that these mathematical and physical 
properties go beyond mere coincidences.
For both the vacuum and non-vacuum cases, Hawking's quasilocal
mass expression agrees with the canonical {\sc qle} results
(\ref{result_one}) and (\ref{result_two}) up to and including
the first non-trivial order in the affine radius. We find this
result rather striking in light of the fact that the Hawking
mass has no apparent connection with the gravitational action or
Hamiltonian. We show that the non-vacuum result (\ref{result_one})
has the expected form based on the results of
Newtonian potential theory.

Compare the small-sphere limit considered here with
the large-sphere limit when $S(r)$ tends to
a cut of $J^{+}$, in which case we know
that the choice of Euclidean reference yields
agreement between $E$ and the {\sc tbs}
energy.\cite{BrownLauYork} In both limiting cases,
the 2-surface $S(r)$ of interest is a cut, level in an
affine radius $r$, of an outgoing congruence of null
geodesics. There is, however, a crucial distinction to
be made. In the small-sphere case, $S(r)$ arises as the
cut of a genuine lightcone, while in the large-sphere
case this is generally not true. We find it remarkable
that this distinction can be mirrored in the choice of
zero-points. To grasp this point, consider first the
Euclidean reference for either limit. This reference
involves an {\em isometric} embedding of $S(r)$ into
a flat Euclidean 3-space $E^{3}$. Now, in both limiting
scenarios $S(r)$ is, in general, slightly distorted
from perfect roundness. Therefore, with $E^{3}$
viewed in turn as an inertial slice of Minkowski space
$\referenceM$, the enveloped $S(r) \subset E^{3} \subset
\referenceM$ cannot be the cut of a genuine lightcone
of $\referenceM$. (Technically, in this case the
outward null congruence associated with $S(r)$ is not
shear-free, but the lightcones of $\referenceM$ are
shear-free.) Therefore, in the general large-sphere
scenario, neither the $S(r)$ embedding into the physical
spacetime $M$ nor its embedding into the reference
spacetime $\referenceM$ corresponds to a lightcone
embedding [that is, in neither case is $S(r)$ the cut
of a lightcone]. Of course, for the small-sphere limit
we may employ either the Euclidean reference or the
lightcone reference.
Of these two choices, the lightcone employs flat
{\em spacetime} to put the reference space on an equal
footing with the small sphere construction in the physical
spacetime. To define the lightcone reference,
we first isometrically embed
$S(r)$ into the lightcone $\referenceN_{q}$ of a point
$q \in \referenceM$, and then select a certain 3-surface
$\referenceSigma \subset \referenceM$ spanning $S(r)$ [so
that $S(r) = \referenceSigma \bigcap \referenceN_{q}$].
The details of this construction, eventually leading to
the expression (\ref{lcmeancurvature}), are found in
Subsection 2.A.
Now, having defined the lightcone reference (tailored to
the small sphere limit) and found the resulting
closed-form expression (\ref{lcmeancurvature}),
we may now invert the question, asking whether or not
a lightcone-reference $k |^{\scriptscriptstyle {\rm ref}}$
defines a {\sc qle}
(\ref{BrownYorkqle}) possessing a correct
large-sphere limit towards $J^{+}$. We discuss this issue
in Appendix A.

The organization of this paper is as follows. In Section 1
we lay the foundations for our examination of the
small-sphere limit. We describe the geometry
of the limit in Subsection 1.A, fix some general conventions in
Subsection 1.B, and make some general observations
concerning the embedding of a $2$-surface in Minkowski
spacetime in Subsection 1.C. No choice of energy zero-point is
made in Section 1, although the results of
Subsection 1.C are used in the subsequent
sections to construct zero-points. In Section 2 we study
the small-sphere limit, subject to the choice of lightcone
reference, and derive the results (\ref{result_one}) and
(\ref{result_two}). In Section 3 we discuss the relationships
between the main results (\ref{result_one}), (\ref{result_two})
and results from Newtonian
potential theory.  Appendix A contains discussions of the small-sphere
limit subject
to the choice of Euclidean reference, and the large-sphere limit
towards $J^{+}$ subject to the choice of lightcone reference.
Throughout our analysis, we use the Newman-Penrose ({\sc np})
formalism\cite{NewmanTod,PenroseRindler}, with which
we assume the reader is familiar. In Appendix B we collect
various conventions and results associated with the {\sc np}
formalism which are used in the main parts of the paper.


\section{Preliminaries}

\subsection{Geometry of the limit construction}

Choose a generic spacetime point $p \in M$, as well as a
unit, future-pointing, timelike vector $u^{\mu}$ which lies
in $T_{p}(M)$, the tangent space at $p$. We may think of
$u^{\mu}$ as the instantaneous four-velocity of an Eulerian
observer at $p$. Let $N_{p} \subset M$ represent the future
lightcone generated by the null geodesics emanating from
$p$. Label the generators of $N_{p}$ by coordinates
$(\theta,\phi)$, or equivalently by $(\zeta,\bar{\zeta})$,
where $\zeta := e^{{\rm i}\phi}\cot(\theta/2)$ is the
stereographic coordinate. Further, choose the affine
parameter $r$ along the generators of the lightcone which
at the point $p$ satisfies the following conditions:
$r = 0$ and the null tangent
$l^{\mu} := (\partial/\partial r)^{\mu}$ to the affinely
parameterized generators obeys $l^{\mu} u_{\mu} = - 1$.
By the term {\em small sphere} we shall mean a 2-surface $S(r)
\subset N_{p}$ level in the coordinate $r$. Provided that we
restrict our attention to small enough values of $r$, we need
not be troubled by conjugate points and each 2-surface $S(r)$
will be only slightly distorted from perfect roundness. On our
lightcone $N_{p}$ we construct a null tetrad $\{l^{\mu},
n^{\mu}, m^{\mu}, \bar{m}{}^{\mu}\}$ as follows. We take
$n^{\mu}$ as the inward null normal to each $S(r)$ (normalized
so that $l^{\mu}n_{\mu} = -1$), and assume that the complex
space leg $m^{\mu}$ points everywhere tangent to each $S(r)$.
Further, via enforcement of the condition
   $\bar{m}^{\lambda} l^{\mu} \nabla_{\mu} m_{\lambda} = 0$,
we remove the freedom to perform $r$-dependent rotations of the
complex dyad. Together with the geodetic property of $l^{\mu}$,
this implies that the spin coefficient $\varepsilon = 0$
[cf.~Eq.~(\ref{spincoefficients}a)].

We also consider a standard pseudo-orthonormal
tetrad
$\{u^{\mu} =: e_{\bot}{}^{\mu}, e_{\rm x}{}^{\mu},
e_{\rm y}{}^{\mu}, e_{\rm z}{}^{\mu}\}$ at the point $p$,
in terms of which we have the following expansions at $p$:
\begin{eqnarray}
l^{\mu} & = &
       u^{\mu} + \sin\theta\cos\phi{\,} e_{\rm x}{}^{\mu}
     + \sin\theta\sin\phi{\,} e_{\rm y}{}^{\mu}
     + \cos\theta {\,}e_{\rm z}{}^{\mu}
\label{triad} \eqnum{\ref{triad}a}\\
n^{\mu} & = &
       {\textstyle \frac{1}{2}}\left[u^{\mu}
     - \left( \sin\theta\cos\phi{\,} e_{\rm x}{}^{\mu}
     + \sin\theta\sin\phi {\,}e_{\rm y}{}^{\mu}
     + \cos\theta{\,} e_{\rm z}{}^{\mu}
\right)\right]{\,} .
\eqnum{\ref{triad}b}
\addtocounter{equation}{1}
\end{eqnarray}
Capital Latin letters, e.~g.~$A,B,C,\cdots$, denote
pseudo-orthonormal tetrad indices and run over
the values $(\bot,{\rm x},{\rm y},{\rm z})$.
The expansions $l^{\mu} = l^{A} e_{A}{}^{\mu}$ and
$n^{\mu} = n^{A} e_{A}{}^{\mu}$
in Eq.~(\ref{triad}) show that the components $l^{A}$
and $n^{A}$ are essentially the first four spherical
harmonics. Standard orthogonality properties of
the spherical harmonics then yield the following
identities:
\begin{eqnarray}
\int\! {\rm d}\!\left\langle\Omega\right\rangle
       l^{A} l^{B}
& = &
       {\textstyle \frac{4}{3}}  u^{A} u^{B}
     + {\textstyle \frac{1}{3}}  g^{AB}
\eqnum{\ref{averages}a} \\
\int\! {\rm d}\!\left\langle\Omega\right\rangle
       l^{A} n^{B}
&  = &
       {\textstyle \frac{1}{3}}  u^{A} u^{B}
     - {\textstyle \frac{1}{6}}  g^{AB}{\,} ,
\label{averages} \eqnum{\ref{averages}b} \\
\int\! {\rm d}\!\left\langle\Omega\right\rangle
       l^{A} l^{B} l^{C} l^{D}
& = &  {\textstyle \frac{1}{5}}\left[
         16 u^{A} u^{B} u^{C} u^{D}
       + 12 u^{(A} u^{B} g^{CD)}
       + g^{(AB} g^{CD)} \right] {\,} ,
\eqnum{\ref{averages}c}
\addtocounter{equation}{1}
\end{eqnarray}
which prove quite useful for reducing many of the
integral expressions encountered below.
Here we make use of the convenient notation
${\rm d}\!\left\langle\Omega\right\rangle :=
(4\pi)^{-1}{\rm d}\Omega$, where ${\rm d}\Omega$
is the area element of a unit-radius round sphere and
the integrations in Eq.~(\ref{averages}) are over the
unit-radius round sphere.

\subsection{Physical and reference energy surface densities}

Let us now define the {\em physical energy surface density}
$(8\pi)^{-1}k$, whose proper surface integral is the
total unreferenced quasilocal energy. Our construction
requires that we fix a 3-dimensional hypersurface $\Sigma$
spanning $S(r)$, or, more precisely, an equivalence class of
spanning 3-surfaces determined by the choice of a unit,
future-pointing, timelike vector $u^{\mu}$ on (and pointing
orthogonal to) $S(r)$. We choose
\begin{equation}
      u^{\mu}
   := {\textstyle \frac{1}{2}}l^{\mu}
    + n^{\mu} {\,},
\end{equation}
which at $p$ agrees with the four-velocity introduced in the
last paragraph. In terms of the convergences $\rho$ and
$- \mu$ of the null normals defined in the appendix
Eqs.~(\ref{spincoefficients}i,k), the mean curvature of
$S(r)$ as embedded in $\Sigma$ is given by
\begin{equation}
       k = 2 \mu + \rho
{\,} .
\label{kequals_2mu+rho}
\end{equation}
We shall write
\begin{equation}
       E |^{\rm {\scriptscriptstyle phy}} := (8\pi)^{-1}
       \int_{S(r)} {\rm d}S k
\label{E_phy}
\end{equation}
for the unreferenced {\sc qle} associated with the
physical slice $\Sigma$ in spacetime $M$.
Likewise, we introduce the
{\em reference energy surface density}
$(8\pi)^{-1} k |^{\scriptscriptstyle {\rm ref}}$, and
with it define the reference contribution to the {\sc qle},
\begin{equation}
       E |^{\rm {\scriptscriptstyle ref}} := (8\pi)^{-1}
       \int_{S(r)} {\rm d}S k
       |^{\rm {\scriptscriptstyle ref}} {\,}.
\label{E_ref}
\end{equation}
As mentioned, $E |^{\rm {\scriptscriptstyle ref}}$ is a
functional solely of the intrinsic geometry of $S(r)$,
although as yet we have made no definite choice for
this functional. In Section 2 we choose the specific
functional stemming from the lightcone reference, while
in Section 3 we choose the one stemming from the Euclidean
reference. The difference
\begin{equation}
       E =
       E |^{\rm {\scriptscriptstyle phy}} -
       E |^{\rm {\scriptscriptstyle ref}}
\label{totalqle}
\end{equation}
is the total referenced {\sc qle} (\ref{BrownYorkqle}).

\subsection{The geometry of 2-surfaces in Minkowski
spacetime}

This subsection collects a few basic results concerning
the reference embedding $S(r) \subset \referenceM$. We
shall use these results later when constructing a particular
reference energy surface density. For notational convenience
here and in what follows, we often use a sans serif
notation for objects associated with Minkowski spacetime
$\referenceM$, and we may write $\referencek$ in place of
$k |^{\rm {\scriptscriptstyle ref}}$. We
restrict our attention to {\em Minkowski-spacetime references}.
That is to say, the reference energy surface density
$(8\pi)^{-1}\referencek$ is determined via an
auxiliary isometric embedding of the 2-surface $S(r)$
into a 3-dimensional hypersurface $\referenceSigma$ which
is itself contained in Minkowski spacetime
$\referenceM$.  Physically, this means that we
{\em assign} the zero value of the energy to that portion of
the slice $\referenceSigma$ contained within $S(r)$.

At the end of the day, our expressions for $\referencek$
depend only on the $S(r)$ 2-metric $\sigma_{ab}$
(with $a,b,c, \cdots$ as $S(r)$ indices), and, therefore,
there is technically no need to consider the reference spacetime.
Nevertheless, in order to motivate and derive our choices,
we begin with the chain of inclusions $S(r) \subset
\referenceSigma \subset \referenceM$ and an
overall spacetime point of view associated with it. Let us
collect a few definitions. Take $\referenceu^{\mu}$ as
the unit future-pointing normal of $\referenceSigma$ as
embedded in $\referenceM$. Let $\referencek_{ab}$ be the
extrinsic curvature tensor associated with the spacelike,
outward-pointing, unit normal of $S(r)$ as embedded in
$\referenceSigma$.
Denote by $\referenceK_{ij}$ the
extrinsic curvature tensor
(with $i,j,k, \cdots$ as $\referenceSigma$ indices)
of $\referenceSigma$ associated with $\referenceu^{\mu}$.
Projection into $S(r)$ of all of
the free indices of $\referenceK_{ij}$ defines an extrinsic
curvature tensor $\referencel{}_{ab}$ on $S(r)$ associated
with $\referenceu{}^{\mu}$. Tangential-normal projection
with respect to $S(r)$ of $\referenceK_{ij}$ defines a covector
$\referenceA_{b}$ on $S(r)$. If $S(r)$
is to arise as a 2-surface
in $\referenceM$, then along with its intrinsic metric
$\sigma_{ab}$ the triple $\{\referencek_{ab},
\referencel_{ab}, \referenceA_{b}\}$ of extrinsic data must
obey certain constraints. These are integrability
criteria relating the intrinsic and extrinsic data of $S(r)$
to the (vanishing) components of the $\referenceM$ Riemann
tensor. Among the constraints for $S(r) \subset \referenceM$
are the following:\footnote{As our use of power series in
$r$ is widespread in this paper, we use parenthesis when
a variable, say ${\cal R}$, is raised to a power. Hence,
$({\cal R}){}^{2} = {\cal R} {\cal R}$ and $({\cal R})^{-1}
= 1/{\cal R}$, while ${\cal R}^{2}$ always denotes the
${\rm O}(r^{2})$ coefficient in the expansion ${\cal R}
= \cdots + r^{2} {\cal R}^{2} + \cdots$.
The only exception to this rule will be the radius $r$ itself.
As there is no possibilty for confusion, we use, for example,
$r^{-1}$ to mean $1/r$.}
\begin{eqnarray}
       (\referencek){}^{2}
     - \referencek{}^{ab}\referencek{}_{ab} -
       (\referencel){}^{2}
     + \referencel{}^{ab}
       \referencel{}_{ab} - {\cal R} & = & 0
\label{embeddingconstraints}
\eqnum{\ref{embeddingconstraints}a} \\
       \referencek_{:b} - \referencek^{a}{}_{b:a}
     + \referenceA_{b} \referencel
     - \referenceA^{a} \referencel_{ab} & = & 0 {\,}.
\eqnum{\ref{embeddingconstraints}b}
\addtocounter{equation}{1}
\end{eqnarray}
Here ${\cal R}$ is the scalar curvature of $S(r)$, and
the colon denotes covariant differentiation intrinsic to
$S(r)$. We shall not have need to consider the other embedding
constraints.

\section{Small-sphere limit}

Throughout this section, the term
$k |^{\scriptscriptstyle {\rm ref}}$ stands for the
explicit expression (\ref{lcmeancurvature}) given in
the introduction, and the functional (\ref{E_ref})
is fixed accordingly.

\subsection{Lightcone reference}
To construct the lightcone reference (see the figure
at the end of this paper), consider
the notations of Subsection 1.C and assume that $S(r)$ is a
cut of a genuine lightcone $\referenceN_{q}$ belonging
to a point $q \in \referenceM$. In this case, the
geodesic congruence $\referenceN_{q}$ is sheer-free,
which means that the complex shear
$\referencesigma := \referencek_{mm} + \referencel_{mm}$
of $\referenceN_{q}$ vanishes. Here
$\referencek_{mm} := \referencek_{ab} m^{a} m^{b}$ and
$\referencel_{mm} := \referencel_{ab} m^{a} m^{b}$
are complex components with respect to the $S(r)$ dyad
$\{m^{a},\bar{m}{}^{a}\}$, respectively capturing
the trace-free pieces of $\referencek_{ab}$ and
$\referencel_{ab}$.
The vanishing of $\referencesigma$ thus implies that the
trace-free piece of $\referencek_{ab}$ equals minus the
trace-free piece of $\referencel_{ab}$; whence
Eq.~(\ref{embeddingconstraints}a) becomes
\begin{equation}
       (\referencek){}^{2}
     - (\referencel){}^{2}
     - 2 {\cal R} = 0 {\,} .
\label{tracesgone}
\end{equation}
We have no need to consider
Eq.~(\ref{embeddingconstraints}b) in this
subsection.\footnote{For
               the lightcone reference we construct in this
               subsection,
               Eq.~(\ref{embeddingconstraints}b) and the
               other embedding constraints not appearing
               above would
               be differential equations determining the
               remaining
               $S(r)$ extrinsic data from $\sigma_{ab}$ and our
               choices for $\referencel$ and $\referencek$.}
We define the convergence $\referencerho
         := {\textstyle \frac{1}{2}}
            \left( \referencek
          - \referencel\right)$
of the outward null normal to $S(r) \subset \referenceN_{q}$,
and with it rewrite Eq.~(\ref{tracesgone}) as
\begin{equation}
       (\referencek){}^{2}
     - \left(\referencek - 2 \referencerho\right){}^{2}
     - 2 {\cal R} = 0 {\,} .
\end{equation}
Now, we are going view $\referencerho
= \referencerho(\zeta,\bar{\zeta};r)$ as some specified
function. Specification of $\referencerho$
(indirectly) fixes a slice
$\referenceSigma$ (or, more precisely, an equivalence class
of slices) spanning $S(r)$, which in turn defines
$\referencek$. With the last equation and some algebra,
we find
\begin{equation}
       \referencek
     = {\textstyle \frac{1}{2}}
       (\referencerho){}^{-1} {\cal R}
     + \referencerho
\label{lightcone_k}
\end{equation}
as the expression for $\referencek$ in terms of the
function $\referencerho$ which is not yet specified. Now, plugging both
the radial expansion (\ref{Gausscurvature}) for ${\cal R}$ and
some radial expansion for $\referencerho$ into
Eq.~(\ref{lightcone_k}), we obtain a radial expansion for
$\referencek$, concerning which we make the following crucial
observations. First (relevant for the non-vacuum case),
a reference convergence of the form
$\referencerho = - r^{-1} + {\rm O}(r)$ determines an
expansion for $\referencek$ up to and including
${\rm O}(r^{2})$ (which is actually higher than the order
needed to get the non-vacuum limit).
Second (relevant for the vacuum case), a reference
convergence of the form
$\referencerho = - r^{-1} + {\rm O}(r^{2})$ determines
an expansion for $\referencek$ up to and including
${\rm O}(r^{3})$. Since we wish our final expression for
$\referencek$ to serve as a proper reference term, we
demand that $\referencerho =
\referencerho(\zeta,\bar{\zeta};r)$ be built purely from
the 2-metric $\sigma_{ab}$ of $S(r)$ (so that $\referencek$
is). This restriction alone hardly fixes the choice of
$\referencerho$ (and $\referencek$). However, for both
scenarios of interest the geometry of the embedding of
$S(r)$ into the {\em physical spacetime} $M$ suggests a
natural choice.

Let us write down our choice for $\referencerho$,
verify that it leads to Eq.~(\ref{lcmeancurvature}), and 
finally discuss why it is physically meaningful. We pick
\begin{equation}
      \referencerho
  = - \sqrt{ {\textstyle \frac{1}{2}} {\cal R} }
      \left[1 + {\textstyle \frac{1}{3}}
      ({\cal R})^{-1}\bar{\thop}\thop\log
      {\cal R}\right] {\,} ,
\label{refrhochoice}
\end{equation}
where $\thop$ is the full ``eth'' operator on $S(r)$
[cf.~Eqs.~(\ref{vacuumthop}) and (\ref{nonvacuumthop})]
and on weight-zero scalars the $S(r)$ Laplacian is
$\Delta = 2\bar{\thop} \thop$.
With this choice for $\referencerho$ (and hence this
choice for $\referenceSigma$), we obtain from
Eq.~(\ref{lightcone_k}) the same closed-form
expression,
\begin{equation}
       \referencek
     = - \sqrt{ {\textstyle \frac{1}{2}}
       {\cal R}}
       \left\{1 + {\textstyle \frac{1}{3}}
       ({\cal R})^{-1}\bar{\thop}\thop\log
       {\cal R} + \left[1 + {\textstyle \frac{1}{3}}
       ({\cal R})^{-1}\bar{\thop}\thop\log
       {\cal R}\right]\!{}^{-1}\right\}
{\,} ,
\label{lightconek_ref}
\end{equation}
as given in Eq.~(\ref{lcmeancurvature}) for $8\pi$ times
the reference energy surface density. Now, let us argue that
Eq.~(\ref{lightconek_ref}) defines a physically sensible
reference surface density for the small-sphere limit.
Using Eqs.~(\ref{vacuumthop}) and (\ref{nonvacuumthop}) 
for the $\thop$ operator and the radial expansion
(\ref{Gausscurvature}) for ${\cal R}$, we find (whether in
vacuo or not) that (\ref{refrhochoice}) satisfies
\begin{equation}
        \referencerho
    = - r^{-1} - {\textstyle \frac{1}{4}} r
      ({\cal R}^{0}  + {\textstyle \frac{1}{3}} \thopbarnaught
      \thopnaught {\cal R}^{0})
      + {\rm O}(r^{2})
{\,} ,
\label{serifrminus}
\end{equation}
where $\thopnaught$ is the unit-radius 
round-sphere ``eth'' operator.
Hence, we have at least $\varrho = - r^{-1} + {\rm O}(r)$
for the reference convergence. To start with, consider the
non-vacuum case and glance at the expansion
(\ref{nonvacexpns}a) for the convergence $\rho$ associated
with the physical embedding $S(r) \subset N_{p}$. We see
that through ${\rm O}(r^{0})$ our choice for $\referencerho$
agrees with the physical $\rho$. Now turn to
the vacuum case and notice that the expansion (\ref{vacexpns}a)
for the convergence $\rho$ associated with the physical
embedding obeys $\rho = - r^{-1} + {\rm O}(r^{3})$. Combining
Eq.~(\ref{serifrminus}) with
the vacuum identity\footnote{One proves the identity as
                           follows. First, as we do in
                           the appendix, calculate the
                           vacuum-case ${\rm O}(r^{0})$
                           piece of the $S(r)$ curvature
                           scalar. One finds ${\cal R}^{0}
                           = - 4(\Psi^{0}_{2}
                           + \bar{\Psi}{}^{0}_{2})$
                           [cf.~Eqs.(\ref{vaccurv}a)
                           and (\ref{zerobianchi}d)].
                           But, as shown in the final part
                           of the appendix, the radial Bianchi
                           identities
                           [cf.~Eq.~(\ref{zerobianchi}d,e,f,g)]
                           imply that
                           $\thopnaught\thopbarnaught
                           \Psi^{0}_{2} =
                           - 3 \Psi^{0}_{2}$ which
                           establishes the result.}
\begin{equation}
\thopnaught\thopbarnaught {\cal R}^{0} = - 3 {\cal R}^{0}
{\,} ,
\label{identityforGauss}
\end{equation}
we have $\varrho = -r^{-1} + {\rm O}(r^{2})$; and,
therefore, now through ${\rm O}(r^{1})$, our choice for
$\referencerho$ agrees with the physical $\rho$.
To sum up, we can state that, whether in vacuo or not,
our choice (\ref{refrhochoice}) for $\referencerho$
determines an embedding of $S(r)$ into $\referenceN_{q}
\subset \referenceM$ which would seem closely related
both intrinsically and {\em extrinsically} to the
physical embedding of $S(r)$ into $N_{p} \subset M$. Let
us now put aside the issue of motivation, and simply expand our
choice (\ref{lightconek_ref}) for $\referencek$ in powers
of $r$. Since $\referencerho = -r^{-1} + {\rm O}(r)$ in
non-vacuum, from Eq.~(\ref{lightcone_k}) we get
\begin{equation}
      k |^{\scriptscriptstyle {\rm ref}} =: \referencek =
     - 2 r^{-1} - {\textstyle \frac{1}{2}}
       r {\cal R}^{0}
     + {\rm O}(r^{2})
\label{nonvack_ref}
\end{equation}
for the non-vacuum
radial expansion of Eq.~(\ref{lightconek_ref}).
Likewise, since $\referencerho = - r^{-1} + {\rm O}(r^{2})$ in
vacuo, from Eq.~(\ref{lightcone_k}) we now get
\begin{equation}
     k |^{\scriptscriptstyle {\rm ref}} =:  \referencek =
     - 2 r^{-1}
     - {\textstyle \frac{1}{2}} r {\cal R}^{0}
     - {\textstyle \frac{1}{2}} r^{2} {\cal R}^{1}
     - {\textstyle \frac{1}{2}} r^{3} {\cal R}^{2}
     + {\rm O}(r^{4}) {\,}
\label{basicvack_ref}
\end{equation}
for the vacuum radial expansion of
Eq.~(\ref{lightconek_ref}).

\subsection{Non-vacuum limit}

Turning our attention to the non-vacuum scenario, we
begin by using the appendix Eqs.~(\ref{nonvacexpns}a,f)
for the convergences $-\mu$ and $\rho$ of the null
normals to determine the following expansion for
$8\pi$ times the physical energy surface
density: [cf.~Eq.~(\ref{kequals_2mu+rho})]
\begin{equation}
       k =
     - 2 r^{-1} + r\left[
       \Psi^{0}_{2}
     + \bar{\Psi}{}^{0}_{2}
     + 2 \Lambda{}^{0}
     + {\textstyle \frac{2}{3}}\Phi^{0}_{11}\right]
     + {\rm O}(r^{2}) {\,},
\label{matter_k}
\end{equation}
Here, $\Psi^{0}_{2}$, $\Phi^{0}_{11}$, and $\Lambda{}^{0}$
are $r \rightarrow 0$ limits of standard curvature terms
from the {\sc np} formalism. Respectively, they are
proportional to a certain null-tetrad component
of the Weyl tensor $C_{\mu\nu\lambda\kappa}$ at $p$,
a certain null-tetrad component of the Ricci tensor
$R_{\mu\nu}$ at $p$, and the Ricci scalar
$R$ at $p$ [cf.~Eqs.~(\ref{npcurvature}c,f,k)].
Because we have adapted our null tetrad to the
lightcone $N_{p}$, the components $\Psi^{0}_{2}$ and
$\Phi^{0}_{11}$ are in fact angle-dependent, i.~e.~they
are functions of $(\zeta,\bar{\zeta})$; however, as
the limit of a scalar function, $\Lambda^{0}$ is not
angle dependent. Next, we substitute both
Eq.~(\ref{matter_k}) and the appendix
Eq.~(\ref{nonvacuumdS}) for the surface element
${\rm d}S$ of $S(r)$ into the basic expression
(\ref{E_phy}) for $E |^{\rm {\scriptscriptstyle phy}}$,
and expand out the result in order to obtain
\begin{equation}
       E |^{\rm {\scriptscriptstyle phy}} =
     - r + r^{3} \int\! {\rm d}\!
       \left\langle\Omega\right\rangle
       \left[
       {\textstyle \frac{1}{2}}(\Psi^{0}_{2}
     + \bar{\Psi}{}^{0}_{2})
     + \Lambda{}^{0}
     + {\textstyle \frac{1}{3}}\Phi^{0}_{00}
     + {\textstyle \frac{1}{3}}\Phi^{0}_{11}\right]
     + {\rm O}(r^{4}) {\,}.
\label{anotherE_phy}
\end{equation}
Like before, $\Phi^{0}_{00}$, an angle-dependent term,
is proportional to a certain null-tetrad component
(\ref{npcurvature}g) of the Ricci tensor at $p$.
We can perform
the integrations remaining in Eq.~(\ref{anotherE_phy}).
The spherical average of the real part of $\Psi^{0}_{2}$
vanishes identically. Indeed, the $r \rightarrow 0$ limit
of Eq.~(\ref{npcurvature}c) shows that
    $\Psi^{0}_{2} + \bar{\Psi}{}^{0}_{2} =
     (l^{A} n^{B} l^{C} n^{D} C_{ABCD})|_{p}$, and by
symmetry the average of $l^{A} n^{B} l^{C} n^{D}$ must
yield terms either proportional to $u^{A} u^{B} u^{C}
u^{D}$ or containing $g^{AB}$. Therefore, that the
average in question vanishes follows from the
index symmetries and trace-free character of the
Weyl tensor.
For the next integration we can straightaway write
$\int\!{\rm d}\!\left\langle\Omega\right\rangle
\Lambda{}^{0} = \frac{1}{24} R |_{p}$. To
evaluate the unit-sphere averages of the other
terms in the integrand, we must use the identities
(\ref{averages}a,b). The $r \rightarrow 0$ of
Eq.~(\ref{npcurvature}g) is
$\Phi^{0}_{00} = {\textstyle \frac{1}{2}}
[l^{A}l^{B} R_{AB}]|_{p}$, which shows that
the average of $\Phi^{0}_{00}$ is readily
obtained with Eq.~(\ref{averages}a). By
rewriting the $r \rightarrow 0$ limit of
Eq.~(\ref{npcurvature}k)
as $\Phi^{0}_{11} = [\frac{1}{2} l^{A} n^{B} R_{AB}
+ \frac{1}{8} R]|_{p}$, we likewise obtain the average of
$\Phi^{0}_{11}$ with Eq.~(\ref{averages}b). Adding together
the individual results for these integrations, we
find
\begin{equation}
       E |^{\rm {\scriptscriptstyle phy}} =
     - r + {\textstyle \frac{1}{18}} r^{3}
       \left.\left[5 R_{\mu\nu} u^{\mu} u^{\nu}
     + 2 R\right]\right|_{p} + {\rm O}(r^{4})
\label{E_physeries}
\end{equation}
as the final limiting expression for the
unreferenced {\sc qle}.

To obtain an expansion analogous to Eq.~(\ref{E_physeries})
for $E |^{\rm {\scriptscriptstyle ref}}$, we must first
explicitly compute the radial expansion for $8\pi$ times
the reference energy density. Putting together
Eqs.~(\ref{nonvack_ref}) and (\ref{nonvaccurv_0}), we get
\begin{equation}
       k |^{\rm {\scriptscriptstyle ref}} =
     - 2 r^{-1} +
       {\textstyle \frac{1}{3}}r
       \left[2\thopbarnaught \Psi^{0}_{1} + 2\thopnaught
       \bar{\Psi}{}^{0}_{1}
     - \thopnaught \Phi^{0}_{10}
     - \thopbarnaught \bar{\Phi}{}^{0}_{10}
     - \Phi^{0}_{00}\right] + {\rm O}(r^{2}) {\,}.
\label{finalk_ref}
\end{equation}
Now, substitute Eqs.~(\ref{finalk_ref}) and (\ref{nonvacuumdS})
into Eq.~(\ref{E_ref}), do some algebra, and perform
a few simple integrations\footnote{Here and in what
follows, ``simple integrations'' refer to the following
version (and its complex conjugate) of the divergence
theorem. Suppose that {\sc sw}$(f) = -1$ (with {\sc sw}
denoting {\em spin weight}). Then with $\thopnaught$
representing the ``eth'' operator on the unit-radius
round sphere, we have
$\int\!{\rm d}\!\left\langle\Omega\right\rangle
\thopnaught f = 0$.}
in order to reach
\begin{equation}
       E |^{\rm {\scriptscriptstyle ref}} =
     - r + {\textstyle \frac{1}{6}} r^{3}
       \int\!{\rm d}\!\left\langle\Omega\right\rangle
       \Phi^{0}_{00} + {\rm O}(r^{4}) {\,}.
\end{equation}
Like before, we use Eq.~(\ref{averages}a) to compute
the remaining integral, thereby finding the desired
result:
\begin{equation}
       E |^{\rm {\scriptscriptstyle ref}} =
     - r + {\textstyle \frac{1}{36}} r^{3}
       \left.\left[4R_{\mu\nu} u^{\mu} u^{\nu}
    +  R\right]\right|_{p} + {\rm O}(r^{4}) {\,},
\label{E_refseries}
\end{equation}
which should be compared with Eq.~(\ref{E_physeries}).

We may now easily obtain the Introduction's result
(\ref{result_one}) for
the non-vacuum limit. Indeed, combination of
Eqs.~(\ref{E_physeries}) and (\ref{E_refseries}) yields
the following result for the total {\sc qle} (\ref{totalqle}):
\begin{equation}
       E = {\textstyle \frac{1}{6}} r^{3}
       \left.\left[G_{\mu\nu} u^{\mu} u^{\nu}\right]
       \right|_{p}
     + {\rm O}(r^{4}) {\,}.
\end{equation}
Notably, this result for $E$ is valid whether or not the
field equations hold, i.~e.~it is a geometric identity.
However, with Einstein's equations
$G_{\mu\nu} = 8\pi T_{\mu\nu}$
we immediately arrive at the Introduction's
result (\ref{result_one}).

\subsection{Vacuum limit}

The derivation of the vacuum limit proceeds along the same
lines as those just considered for the non-vacuum limit.
With the appendix Eqs.~(\ref{vacexpns}a,f) for the vacuum-case
convergences $-\mu$ and $\rho$, we compute the radial expansion
for the physical energy surface
density, [cf.~Eq.~(\ref{kequals_2mu+rho})]
\begin{eqnarray}
k        & = &
            - 2 r^{-1}
            + {\textstyle \frac{1}{3}} r
              \left[\thopnaught \bar{\Psi}^{0}_{1}
            + \thopbarnaught \Psi^{0}_{1}\right]
            + {\textstyle \frac{1}{6}}r^{2}\left[
              \thopnaught \bar{\Psi}^{1}_{1}
            + \thopbarnaught \Psi^{1}_{1}\right]
\nonumber     \\
         &   &
            + r^{3} \left[{\textstyle \frac{1}{180}}
              \Psi^{0}_{0}\bar{\Psi}{}^{0}_{0}
            - {\textstyle \frac{1}{20}}
              \thopnaught \left(
              \bar{\Psi}{}^{0}_{0}\Psi^{0}_{1}
            + 4 \bar{\Psi}^{2}_{1}\right)
            - {\textstyle \frac{1}{20}}
              \thopbarnaught \left(\Psi^{0}_{0}
              \bar{\Psi}{}^{0}_{1}
            + 4 \Psi^{2}_{1}\right)
            - {\textstyle \frac{1}{2}}
              {\cal R}^{2}\right]
            + {\rm O}(r^{4}) {\,} .         \label{vack_phy}
\end{eqnarray}
Next, we substitute this expansion along with the expansion
(\ref{vacuumdS}) for the vacuum-case surface element into
Eq.~(\ref{E_phy}), do some algebra, and perform a few simple
integrations. These steps lead to
\begin{equation}
       E |^{\rm {\scriptscriptstyle phy}} = - r +
       r^{5} \int\! {\rm d}\!
       \left\langle\Omega\right\rangle
       \left[{\textstyle \frac{7}{360}}
               \Psi^{0}_{0} \bar{\Psi}{}^{0}_{0}
             + {\textstyle \frac{1}{4}}
               \left({\textstyle \frac{1}{45}}
               \Psi^{0}_{0} \bar{\Psi}{}^{0}_{0}
             - {\cal R}^{2}\right)\right] + {\rm O}(r^{6})
{\,} .
{\,}
\end{equation}
A glance at the explicit expression (\ref{vaccurv}c)
for the coefficient ${\cal R}^{2}$ [of the ${\rm O}(r^{2})$
term in the expansion (\ref{Gausscurvature})
for ${\cal R}$] shows that the unit-sphere
average of the term within the parenthesis above vanishes.
To evaluate the final unit-sphere average, we first note that
the $r \rightarrow 0$ limit of the square of
Eq.~(\ref{npcurvature}a) is
in fact $\Psi^{0}_{0}\bar{\Psi}{}^{0}_{0} = [\frac{1}{4}
l^{A} l^{B} l^{C} l^{D} T_{ABCD}]|_{p}$. This may be checked by
an explicit computation using our definition (\ref{BelRobinson})
of the Bel-Robinson tensor. Evidently
then, we may use Eq.~(\ref{averages}c) to find
\begin{equation}
       E |^{\rm {\scriptscriptstyle phy}} =
       - r  + {\textstyle \frac{7}{450}} r^{5}
              \left.\left[T_{\mu\nu\rho\sigma}
               u^{\mu} u^{\nu} u^{\rho} u^{\sigma}
              \right]\right|_{p} + {\rm O}(r^{6})
\label{vacuumE_phy}
\end{equation}
as the desired limit expression for
$E |^{\rm {\scriptscriptstyle phy}}$ in vacuo.

Turning now to the calculation of
$E |^{\rm {\scriptscriptstyle ref}}$, we first
put together Eqs.~(\ref{basicvack_ref}) and (\ref{vaccurv}),
in order to get the following explicit expansion:
\begin{eqnarray}
k |^{\rm {\scriptscriptstyle ref}}
& = &
            - 2 r^{-1}
            + {\textstyle \frac{2}{3}} r
              \left[\thopnaught \bar{\Psi}^{0}_{1}
            + \thopbarnaught \Psi^{0}_{1}\right] \nonumber \\
& & 
            + {\textstyle \frac{5}{12}}r^{2}\left[
              \thopnaught \bar{\Psi}^{1}_{1}
            + \thopbarnaught \Psi^{1}_{1}\right]
            + {\textstyle \frac{1}{2}} r^{3}\left[
              \left({\textstyle \frac{1}{45}}
              \Psi^{0}_{0}\bar{\Psi}{}^{0}_{0}
            - {\cal R}^{2}\right)
            - {\textstyle \frac{1}{45}}
              \Psi^{0}_{0}\bar{\Psi}{}^{0}_{0}\right]
            + {\rm O}(r^{4})  {\,} .
\label{vack_ref}
\end{eqnarray}
We substitute this expansion as well as the expansion
(\ref{vacuumdS}) into Eq.~(\ref{E_ref}), and
again do some algebra and
a few simple integrations, thereby obtaining
\begin{equation}
       E |^{\rm {\scriptscriptstyle ref}} = - r
       + r^{5} \int\! {\rm d}\!
       \left\langle\Omega\right\rangle
       \left[{\textstyle \frac{1}{180}}
               \Psi^{0}_{0} \bar{\Psi}{}^{0}_{0}
             + {\textstyle \frac{1}{4}}
               \left({\textstyle \frac{1}{45}}
               \Psi^{0}_{0} \bar{\Psi}{}^{0}_{0}
             - {\cal R}^{2}\right)\right] + {\rm O}(r^{6}) {\,}.
\end{equation}
Finally, calculations identical to those just performed for
$E |^{\rm {\scriptscriptstyle phy}}$ establish that
\begin{equation}
       E |^{\rm {\scriptscriptstyle ref}} =
       - r  + {\textstyle \frac{1}{225}} r^{5}
              \left.\left[T_{\mu\nu\rho\sigma}
               u^{\mu} u^{\nu} u^{\rho} u^{\sigma}
              \right]\right|_{p} + {\rm O}(r^{6}) {\,},
\label{vacuumE_ref}
\end{equation}
and the difference of
(\ref{vacuumE_phy}) and (\ref{vacuumE_ref}) immediately
gives the Introduction's result (\ref{result_two}).

\section{Newtonian Potential Theory}
It is interesting to compare our main results with analogous
results from Newtonian potential theory.\footnote{A comparison
between the canonical {\sc qle} with Euclidean subtraction and
the Newtonian gravitational energy is given in
Ref.~\cite{BrownYork}.}
The Newtonian interpretation for the Introduction's result
(\ref{result_one}) is straightforward.
Consider a pressureless ball of fluid of radius $r$ and
constant (volume) mass density $\epsilon$.
The total Newtonian mass
for the ball is $M = \frac{4}{3}\pi r^{3} \epsilon$.
If we identify this radius with the affine radius that appears
in the result (\ref{result_one}), then a further
identification between the Newtonian
mass $M$ and the quasilocal energy $E$
yields the correspondence
$\epsilon = [T_{\mu\nu} u^{\mu} u^{\nu}]|_{p}$. This is a solid
result, fully in accord with the non-existence of ``pure
gravitational field energy" at a point, since only matter energy
contributes in this limit.
The Newtonian analog of the result
(\ref{result_two}) is not clear. Nevertheless, we find the
following result rather interesting.
Consider again the ball of fluid, but now let us compute
the Newtonian gravitational field energy within the ball.
Recall that the energy density of the Newtonian gravitational
field is given by $-\vec\nabla \Phi \cdot \vec\nabla \Phi/(8\pi)$,
where $\vec\nabla$ is the flat-space gradient operator and 
$\Phi$ is the
Newtonian potential. Inserting the appropriate expression for
$\Phi$ into the energy density and integrating over the interior
of the ball, we obtain
\begin{equation}
E_N = -{\textstyle \frac{1}{90}}r^{5}(4\pi \epsilon)^{2}
\end{equation}
for the gravitational potential energy inside the ball.
Comparison of this result with the
small-sphere  result (\ref{result_two}), we see that both expressions
depend on the fifth power of radius and both contain a numerical factor
of
$\frac{1}{90}$. We do not know at this time if the close resemblance
between these results has any physical significance.

\section{Acknowledgments}
For discussions we thank H.~Balasin,
T.~G.~Concannon, N. \'{O} Murchadha, and L.~B.~Szabados.
In particular, SRL thanks L.~B.~Szabados for
comments which led to this investigation and N. \'{O}
Murchadha for discussions concerning the
Newtonian interpretation of the vacuum limit. This work was
begun while SRL visited the Research Institute for Particle
\& Nuclear Physics in Budapest, Hungary,
and SRL wishes to express his appreciation to this
institute both for financial support and a hospitable stay
in Budapest.  This work has been supported by the National
Science Foundation of the USA (NSF grant \# PHY-9413207 to the
University of North Carolina), and
in part by the ``Fonds zur F\"{o}r\-der\-ung
der wis\-sen\-schaft\-lich\-en For\-schung''
in Austria (FWF Project 10.221-PHY).

\appendix
\section{Euclidean Reference and Large-sphere limit}
One can also carry out our analysis of the small-sphere limit
for the choice of Euclidean reference.
To define the Euclidean reference, again consider the
notations of Subsection I.C,
but now assume that $\referenceSigma$
is a flat inertial hyperplane $E^{3}$ in
$\referenceM$. In this case,
$\referencel_{ab} = 0 = \referenceA_{b}$ and the
constraints given in
Eqs.~(\ref{embeddingconstraints}a,b) are simply
\begin{eqnarray}
       (\referencek){}^{2}
     - \referencek{}^{ab} \referencek{}_{ab}
     - {\cal R} = 0
\label{scalarandvector} \eqnum{\ref{scalarandvector}a}\\
       \referencek{}_{:b}
     - \referencek{}^{a}{}_{b:a} = 0 {\,} .
\eqnum{\ref{scalarandvector}b}
\addtocounter{equation}{1}
\end{eqnarray}
To analyze these equations,
it proves useful to split $\referencek{}_{ab}$
into its trace and trace-free pieces. This is readily achieved
by working with the components of $\referencek{}_{ab}$ with
respect to the null dyad $\{m^{a}, \bar{m}{}^{a}\}$. Indeed,
$2\referencek{}_{m\bar{m}} := 2 m^{a} \bar{m}{}^{b}
\referencek{}_{ab}$ and $\referencek{}_{mm} := m^{a} m^{b}
\referencek{}_{ab}$ respectively capture the trace and
trace-free pieces of $\referencek_{ab}$, and in terms of
these quantities
the Eqs.~(\ref{scalarandvector}a,b) become
\begin{eqnarray}
       [\referencek{}_{m\bar{m}}]^{2}
     - \referencek{}_{mm} \referencek{}_{\bar{m}\bar{m}}
     - {\textstyle \frac{1}{2}}{\cal R} & = & 0
\label{scalarandvector2} \eqnum{\ref{scalarandvector2}a}\\
       \thop \referencek{}_{m\bar{m}}
     - \bar{\thop} \referencek{}_{mm} & = & 0 {\,}.
\eqnum{\ref{scalarandvector2}b}
\addtocounter{equation}{1}
\end{eqnarray}
For the case of Euclidean reference, we are unable to obtain a
closed-form expression for $\referencek$ (neither in the
non-vacuum nor vacuum cases). Nevertheless, it is
evident from Eqs.~(\ref{scalarandvector}a,b) that the full
extrinsic curvature tensor $\referencek_{ab}$ (and hence
its trace piece) is determined solely by the intrinsic
metric on $S(r)$. One expects that
Eqs.~(\ref{scalarandvector2}a,b) may be solved for
$\referencek_{ab}$,
provided $S(r)$ is only slightly distorted from perfect
roundness.
For small enough values of $r$, the expansion
(\ref{Gausscurvature}) for ${\cal R}$ assures us that
this is indeed the case. Moreover, the solution
$\referencek_{ab}$ should be unique up to Euclidean
translations and rotations. Lacking a closed-form expression
for $\referencek$, we have obtained radial expansions in
both the non-vacuum and vacuum
scenarios. For both scenarios, we obtain such expansions by
first plugging into Eqs.~(\ref{scalarandvector2}a,b) the
{\em Ans\"{a}tze}
\begin{eqnarray}
       \referencek_{m\bar{m}}
& = &
     - r^{-1}
     + r \referencek^{1}_{m\bar{m}}
     + r^{2} \referencek^{2}_{m\bar{m}}
     + r^{3} \referencek^{3}_{m\bar{m}}
     + {\rm O}(r^{4})
\label{ansatz} \eqnum{\ref{ansatz}a} \\
       \referencek_{mm}
& = &
       r \referencek^{1}_{mm}
     + r^{2} \referencek^{2}_{mm}
     + r^{3} \referencek^{3}_{mm}
     + {\rm O}(r^{4})
{\,} ,
\eqnum{\ref{ansatz}b}
\addtocounter{equation}{1}
\end{eqnarray}
along with appropriate (vacuum or non-vacuum, as the case
may be) expansions
for both $\thop$ and ${\cal R}$. We have performed this calculation,
and find that the choice of Euclidean reference also establishes
(\ref{result_one}) in the non-vacuum case.  However, with the
choice of Euclidean reference, the vacuum small-sphere limit of the
{\sc qle} is not related directly to the Bel-Robinson tensor.

Our closing
comments concern the large-sphere limit, in which
case $S(r)$ tends to a round infinite-radius 2-sphere cut of
$J^{+}$. Thus, we now consider a spacetime
$M$ which is asymptotically flat towards $J^{+}$ and a
corresponding
system of Bondi coordinates $(w,r,\zeta,\bar{\zeta})$.
Here $w$ is a retarded time coordinate, and $r$ is an
affine radius similar to before. Now $S(r)$ arises as a cut,
level in $r$, of an outgoing null hypersurface ${\cal N}$ (in
general not a genuine lightcone), and twice the Gaussian
curvature of $S(r)$ has the following expansion in powers of
inverse $r$:
\begin{equation}
{\cal R} = 2r^{-2} + {\cal R}^{-3} r^{-3} + {\rm O}(r^{-4})
{\,} ,
\end{equation}
where the coefficient ${\cal R}^{-3}$ may be expressed in
terms of the asymptotic shear $\sigma^{0}$ of
${\cal N}$.\cite{BrownLauYork} If we use formula (\ref{tracesgone})
(appropriate for a lightcone embedding) along with the ``rest-frame''
choice $\referencel = 0$, then we have
\begin{equation}
\referencek = -\sqrt{2{\cal R}} {\,} .
\label{simple_reference}
\end{equation}
Although this expression differs from (\ref{lcmeancurvature}), it is
also a lightcone reference. from (\ref{simple_reference}) we find
the following radial expansion for the reference mean curvature:
\begin{equation}
k |^{\scriptscriptstyle {\rm ref}} =
- 2r^{-1} - {\textstyle \frac{1}{2}} r^{-2}
{\cal R}^{-3} + {\rm O}(r^{-4}) {\,} ,
\end{equation}
which agrees through ${\rm O}(r^{-2})$
with the $k |^{\scriptscriptstyle {\rm ref}}$
expansion obtained in Ref.~\cite{BrownLauYork} via the
Euclidean reference. Hence, all of the results found in
Ref.~\cite{BrownLauYork} are also valid for
the this choice of lightcone reference. In particular, the
lightcone-referenced {\sc qle} agrees with the {\sc tbs} energy in a
suitable null limit. Moreover, in the same limit the
``smeared'' version of the {\sc qle}  [which incorporates
a lapse function into the definition (\ref{BrownYorkqle})]
agrees with Geroch's supermomentum
(when the latter is evaluated in a Bondi conformal frame).
See Ref.~\cite{BrownLauYork} for further details.
To differentiate between the lightcone
and Euclidean references in the
large-sphere limit, one could examine
multipole-moment terms (which
arise at higher powers of inverse radius)
for stationary spacetimes.
We hope to return to this issue elsewhere.

\section{Connection and Curvature}
Throughout our discussion, we have taken $(-,+,+,+)$ as the
signature of the spacetime metric. We have also adopted the
index conventions of Ref.~\cite{MTW} for curvature. With
these choices we shall define the standard Newman-Penrose
({\sc np}) connection and curvature coefficients such that
our {\sc np} equations match those listed by Newman and Tod
in the appendix of Ref.~\cite{NewmanTod}. As we use a somewhat
non-standard {\sc np} formalism, let us list some of our
basic conventions explicitly.
Label our null frame as
$\{l^{\mu},n^{\mu},m^{\mu},\bar{m}{}^{\mu}\} =
\{e_{1}{}^{\mu}, e_{2}{}^{\mu}, e_{3}{}^{\mu}, e_{4}{}^{\mu}\}$
and name its associated
connection coefficients as follows:
\begin{eqnarray}
       \varepsilon
     = {\textstyle \frac{1}{2}}(\Gamma_{121} + \Gamma_{431})
     \hspace{1cm} & \kappa = \Gamma_{131} \hspace{1cm} &
       \pi = \Gamma_{421}
\label{spincoefficients} \eqnum{\ref{spincoefficients}a,b,c} \\
       \gamma
     = {\textstyle \frac{1}{2}}(\Gamma_{122} + \Gamma_{432})
     \hspace{1cm} & \tau = \Gamma_{132} \hspace{1cm} &
       \nu = \Gamma_{422}
\eqnum{\ref{spincoefficients}d,e,f} \\
       \beta
     = {\textstyle \frac{1}{2}}(\Gamma_{123} + \Gamma_{433})
     \hspace{1cm} & \sigma = \Gamma_{133} \hspace{1cm} &
       \mu = \Gamma_{423}
\eqnum{\ref{spincoefficients}g,h,i} \\
       \alpha
     = {\textstyle \frac{1}{2}}(\Gamma_{124} + \Gamma_{434})
     \hspace{1cm} & \rho = \Gamma_{134} \hspace{1cm} &
       \lambda = \Gamma_{424}{\,} ,
\eqnum{\ref{spincoefficients}j,k,l}
\addtocounter{equation}{1}
\end{eqnarray}
where, for example, $\Gamma_{134} = l^{\mu} \bar{m}{}^{\nu}
\nabla_{\nu}m_{\mu}$. For the small-sphere limit the null frame
is adapted to the lightcone $N_{p}$ as described in the first
section, and, as a result, the following vanish identically
for our construction:
\begin{equation}
\rho - \bar{\rho} = \mu - \bar{\mu} =
\varepsilon = \kappa = \tau - \bar{\alpha} - \beta
= \pi - \alpha - \bar{\beta} = 0 {\,}.
\label{vanishingGammas}
\end{equation}
As we retain the freedom to perform $r$-independent rotations
of the space leg $m^{\mu}$, we shall use those elements of
the Geroch-Held-Penrose ({\sc ghp})
formalism\cite{PenroseRindler}
pertaining both to spin-weighted scalars and the
``eth'' operator $\thop$ on $S(r)$. If a quantity $Q$
transforms as $Q \rightarrow \exp(2{\rm i}s \chi)Q$ under the
rotation $m^{\mu} \rightarrow \exp(2{\rm i}\chi)m^{\mu}$ (with
$\chi$ independent of $r$), then $Q$ is said to be a
spin-weighted scalar of spin weight $s$ (in symbols,
{\sc sw}$(Q) = s$). As the extents of our null normals
$l^{\mu}$ and $n^{\mu}$ have been fixed once and for all,
we have no need to consider the concept of boost weight.

Next, consider the spacetime Ricci scalar
as well as the components of the
Weyl and Ricci tensors with
respect to the null tetrad. With these define the following
standard pieces of the spacetime
curvature:
\begin{eqnarray}
\Psi_{0} = C_{1313} & \Psi_{1} = C_{1213} & \Psi_{2}
= {\textstyle \frac{1}{2}}(C_{1212} + C_{4312})
\label{npcurvature} \eqnum{\ref{npcurvature}a,b,c} \\
\Psi_{3} = C_{1242} & \Psi_{4} = C_{2424} &
\Lambda  = {\textstyle \frac{1}{24}} R
\eqnum{\ref{npcurvature}d,e,f} \\
\Phi_{00} = {\textstyle \frac{1}{2}} R_{11}
& \Phi_{10} = {\textstyle \frac{1}{2}} R_{14} &
\Phi_{20} = {\textstyle \frac{1}{2}} R_{44}
\eqnum{\ref{npcurvature}g,h,i} \\
\Phi_{01} = {\textstyle \frac{1}{2}} R_{13}
& \hspace{2.45cm}
\Phi_{11} = {\textstyle \frac{1}{4}} (R_{12} + R_{34})
\hspace{1cm} &
\Phi_{21} = {\textstyle \frac{1}{2}} R_{24}
\eqnum{\ref{npcurvature}j,k,l} \\
\Phi_{02} = {\textstyle \frac{1}{2}} R_{33}
& \Phi_{12} = {\textstyle \frac{1}{2}} R_{23} &
\Phi_{22} = {\textstyle \frac{1}{2}} R_{22} {\,} .
\eqnum{\ref{npcurvature}m,n,o}
\addtocounter{equation}{1}
\end{eqnarray}
As mentioned above, with these conventions our {\sc np}
equations are exactly those listed in the appendix of
Ref.~\cite{NewmanTod}; however, we shall consider all possible
simplification of these equations afforded by
Eq.~(\ref{vanishingGammas}) and the use of $\thop$
[cf.~Eqs.~(\ref{nonvacuumthop}) and (\ref{vacuumthop})].
For the small-sphere limit examined in this paper, we
consider the pullbacks to the lightcone $N_{p}$ of
the curvature components (\ref{npcurvature}) and assume that
each pullback may be expanded as a power series in $r$ along
the lightcone. That is, we assume
\begin{eqnarray}
\Psi_{i} = \Psi^{0}_{i} + r \Psi^{1}_{i}
+ r^{2} \Psi^{2}_{i} + \cdots & \hspace{2cm} &
i = 0,1,2,3,4 \label{curvexpns} \eqnum{\ref{curvexpns}a}\\
\Phi_{ij} = \Phi^{0}_{ij} + r \Phi^{1}_{ij}
+ r^{2} \Phi^{2}_{ij} + \cdots
& & i,j = 0,1,2\eqnum{\ref{curvexpns}b} \\
\Lambda = \Lambda^{0} + r \Lambda^{1}
+ r^{2} \Lambda^{2} + \cdots {\,}.
& & \eqnum{\ref{curvexpns}c}
\end{eqnarray}
We shall use these expansions for the curvature components
along with the radial {\sc np} field equations in order to
obtain radial expansions along $N_{p}$
for those spin coefficients
(\ref{spincoefficients}) used in our analysis.
We consider the non-vacuum and vacuum cases separately.

\subsection{Non-vacuum}
Up to the appropriate order in the affine radius, we have
confirmed the non-vacuum asymptotic expansions of the
spin coefficients given by Kelly {\em et al}\cite{KTW} and
by Dougan.\cite{Dougan}
For completeness we recall that this list is
\begin{eqnarray}
\rho    & = &
           - r^{-1} + {\textstyle \frac{1}{3}}r\Phi^{0}_{00}
           + {\rm O}(r^{2})
            \label{nonvacexpns} \eqnum{\ref{nonvacexpns}a} \\
\sigma  & = &
             {\textstyle \frac{1}{3}} r \Psi^{0}_{0}
           + {\rm O}(r^{2})     \eqnum{\ref{nonvacexpns}b} \\
\alpha  & = &
             r^{-1} \alpha^{0}
           + {\textstyle \frac{1}{6}} r\left[\alpha^{0}
             \Phi^{0}_{00} + 2 \Phi^{0}_{10}
           - \bar{\Psi}^{0}_{1}
           - \bar{\alpha}{}^{0}\bar{\Psi}{}^{0}_{0}\right]
           + {\rm O}(r^{2})     \eqnum{\ref{nonvacexpns}c} \\
\beta   & = &
           - r^{-1} \bar{\alpha}{}^{0}
           - {\textstyle \frac{1}{6}} r\left[\bar{\alpha}{}^{0}
             \Phi^{0}_{00} - 3\Psi^{0}_{1}
           - \alpha^{0}\Psi^{0}_{0}\right]
           + {\rm O}(r^{2})     \eqnum{\ref{nonvacexpns}d} \\
\lambda & = &
             {\textstyle \frac{1}{6}}r \left[5 \Phi^{0}_{20}
           + \bar{\Psi}{}^{0}_{0}\right]
           + {\rm O}(r^{2})     \eqnum{\ref{nonvacexpns}e} \\
\mu     & = &
           - {\textstyle \frac{1}{2}} r^{-1}
           + {\textstyle \frac{1}{2}} r \left[ \Psi^{0}_{2}
           + \bar{\Psi}{}^{0}_{2} + 2 \Lambda^{0}
           + {\textstyle \frac{2}{3}}\Phi^{0}_{11}
           - {\textstyle \frac{1}{3}}\Phi^{0}_{00}\right]
           + {\rm O}(r^{2})     \eqnum{\ref{nonvacexpns}f}
             {\,} ,
\addtocounter{equation}{1}
\end{eqnarray}
where $\alpha^{0} \equiv \sqrt{{\textstyle \frac{1}{8}}} \zeta$.
For the scenario at hand the expansions (\ref{nonvacexpns}c,d)
determine the expansions for $\tau$ and $\pi$ up to
${\rm O}(r^{2})$. Actually, we need only know that $\lambda =
{\rm O}(r)$ in order to obtain the ${\rm O}(r)$ coefficient for
$\mu$; however, for completeness we have explicitly given the
${\rm O}(r)$ coefficient for $\lambda$.
As they are not needed in this
paper, we do not list the expansions for $\gamma$ and $\nu$.

\subsection{Vacuum}
For the vacuum scenario, we need some of the required
spin-coefficient expansions out to an order higher than
given in either of Refs.~\cite{KTW,Dougan}. We obtain the
following list:
\begin{eqnarray}
\rho     & = &
            - r^{-1} + {\textstyle \frac{1}{45}} r^{3}
              \Psi^{0}_{0} \bar{\Psi}^{0}_{0}
            + {\rm O}(r^{4})
                 \label{vacexpns} \eqnum{\ref{vacexpns}a} \\
\sigma   & = &
              {\textstyle \frac{1}{3}}r \Psi^{0}_{0}
            + {\textstyle \frac{1}{4}}r^{2}\Psi^{1}_{0}
            + {\textstyle \frac{1}{5}}r^{3}\Psi^{2}_{0}
            + {\rm O}(r^{4})      \eqnum{\ref{vacexpns}b} \\
\alpha   & = &
              r^{-1} \alpha^{0}
            - {\textstyle \frac{1}{6}}r
              \left[\bar{\Psi}{}^{0}_{1} + \bar{\alpha}{}^{0}
              \bar{\Psi}{}^{0}_{0}\right]
            - {\textstyle \frac{1}{12}} r^{2}
              \left[\bar{\Psi}{}^{1}_{1}
            + \bar{\alpha}{}^{0}
              \bar{\Psi}{}^{1}_{0}\right] \nonumber       \\
         &   &
            + {\textstyle \frac{1}{360}}
              r^{3}\left[7 \alpha^{0}
              \Psi^{0}_{0} \bar{\Psi}{}^{0}_{0}
            + 8 \bar{\Psi}{}^{0}_{0}\Psi^{0}_{1}
            - 18 \bar{\alpha}{}^{0} \bar{\Psi}{}^{2}_{0}
            - 3 \thopnaught \bar{\Psi}{}^{2}_{0}\right]
            + {\rm O}(r^{4})       \eqnum{\ref{vacexpns}c} \\
\beta    & = &
            - r^{-1} \bar{\alpha}{}^{0}
            + {\textstyle \frac{1}{6}} r
              \left[3\Psi^{0}_{1}
            + \alpha^{0} \Psi^{0}_{0}\right]
            + {\textstyle \frac{1}{12}} r^{2}
              \left[4\Psi^{1}_{1}
            + \alpha^{0} \Psi^{1}_{0}\right] \nonumber     \\
         &   &
            - {\textstyle \frac{1}{360}} r^{3}
              \left[7 \bar{\alpha}{}^{0}
              \Psi^{0}_{0} \bar{\Psi}{}^{0}_{0}
            - 20 \Psi^{0}_{0}\bar{\Psi}{}^{0}_{1}
            - 18 \alpha^{0} \Psi^{2}_{0}
            - 15 \thopbarnaught \Psi^{2}_{0}\right]
            + {\rm O}(r^{4})        \eqnum{\ref{vacexpns}d} \\
\lambda  & = &
            - {\textstyle \frac{1}{6}}r\bar{\Psi}{}^{0}_{0}
            + {\rm O}(r^{2})
                                     \eqnum{\ref{vacexpns}e} \\
\mu      & = &
            - {\textstyle \frac{1}{2}}r^{-1}
            + {\textstyle \frac{1}{2}} r
              \left[\Psi^{0}_{2} + \bar{\Psi}{}^{0}_{2}\right]
            + {\textstyle \frac{1}{3}} r^{2}
              \left[ \Psi^{1}_{2}
            + \bar{\Psi}{}^{1}_{2}\right]     \nonumber     \\
         &   &
            + r^{3} \left[{\textstyle \frac{1}{360}}
              \Psi^{0}_{0}\bar{\Psi}{}^{0}_{0}
            - {\textstyle \frac{1}{40}}
              \thopnaught \left(
              \bar{\Psi}{}^{0}_{0}\Psi^{0}_{1}
            + 4 \bar{\Psi}^{2}_{1}\right)
            - {\textstyle \frac{1}{40}}
              \thopbarnaught \left(\Psi^{0}_{0}
              \bar{\Psi}{}^{0}_{1}
            + 4 \Psi^{2}_{1}\right)
            - {\textstyle \frac{1}{4}}
              {\cal R}^{2}\right]
            + {\rm O}(r^{4}) {\,} ,
\eqnum{\ref{vacexpns}f}
\addtocounter{equation}{1}
\end{eqnarray}
where the coefficient ${\cal R}^{2}$
[cf.~Eq.~(\ref{Gausscurvature})]
found in $\mu$ is written out explicitly below
in Eq.~(\ref{vaccurv}c).
Also, the unit-sphere ``eth'' operator
$\thopnaught$ is defined below in Eq.~(\ref{ethnaught}).
As before, Eqs.~(\ref{vacexpns}c,d)
determine expansions for
$\tau$ and $\pi$ up to ${\rm O}(r^{4})$,
and we do not need the expansions for $\gamma$ and $\nu$.

The particular form of the ${\rm O}(r^{3})$ coefficient
$\mu^{3}$ in the expansion (\ref{vacexpns}f) for
\begin{equation}
\mu = - {\textstyle \frac{1}{2}} r^{-1} + \cdots +
                     r^{3}\mu^{3} + {\rm O}(r^{4})
{\,}
\label{muexpansion}
\end{equation}
plays a crucial role in our calculation of the {\sc qle}'s
vacuum limit. Therefore, let us sketch how to obtain the
given form of $\mu^{3}$, assuming that we have already
determined both $\mu$ up to and including ${\rm O}(r^{2})$
and the remaining expansions (\ref{vacexpns}) as given.
Rather than making a  straightforward appeal to the {\sc np}
field equations (as we have done to obtain the other spin
coefficients), we shall instead take advantage of the
particular geometry of our construction and derive this
coefficient via the geometric identity \cite{PenroseRindler}
\begin{equation}
K = - \Psi_{2} + \Phi_{11} + \Lambda - \sigma\lambda
    + \mu\rho
{\,} .
\label{complexGauss}
\end{equation}
$K$ is the complex Gauss curvature of $S(r)$, and the
Ricci scalar of $S(r)$ is simply
${\cal R} = 2(K + \bar{K})$. Since we work here in
vacuo and both $\rho$ and $\mu$ are real,
Eq.~(\ref{complexGauss}) implies that
\begin{equation}
{\cal R} = - 2(\Psi_{2} + \bar{\Psi}_{2}) - 2(\sigma\lambda
+ \bar{\sigma} \bar{\lambda} ) + 4\mu\rho
{\,} .
\label{Randmu}
\end{equation}
Now, into this equation insert both Eq.~(\ref{muexpansion})
and the expansions (\ref{vacexpns}a,b,e), and then isolate the
${\rm O}(r^{2})$ piece of the resulting expression, in order
to establish that
\begin{equation}
       \mu^{3}
   = - {\textstyle \frac{1}{2}}
       (\Psi^{2}_{2} + \bar{\Psi}{}^{2}_{2})
     + {\textstyle \frac{2}{45}}\Psi^{0}_{0}
       \bar{\Psi}{}^{0}_{0}
     - {\textstyle \frac{1}{4}}{\cal R}^{2}
{\,} .
\label{muthree}
\end{equation}
To work this relationship into the desired form, we appeal to
the following vacuum {\sc np} Bianchi identity:
\begin{equation}
D \Psi_{2} - \bar{\thop}\Psi_{1} =
- \lambda \Psi_{0} + \pi \Psi_{1} + 3 \rho \Psi_{2}
= 0 {\,} ,
\label{Bianchi}
\end{equation}
where $D = l^{\mu} \nabla_{\mu}$
and the action of $\bar{\thop}$
on $\Psi_{1}$ is defined as the conjugate of
$\thop \bar{\Psi}_{1}$ [with
$\thop$ as in Eq.~(\ref{vacuumthop})
and {\sc sw}$(\bar{\Psi}_{1}) = -1$]. The given form of
Eq.~(\ref{Bianchi}) is specific to the geometry of our
construction, but it can be deduced from the (more general)
appendix Eq.~(A.4c) given in Ref.~\cite{NewmanTod}.
Now, plug the expansions (\ref{curvexpns}) and (\ref{vacexpns})
as well as the
radial expansion for $\bar{\thop}$ into the Bianchi
identity (\ref{Bianchi}), thereby obtaining a
tower of identities (one identity at each power of $r$).
Isolate the particular identity determined at ${\rm O}(r)$
in the tower, and into this equation make repeated
substitutions with (\ref{zerobianchi}a,b).
These steps lead to
\begin{equation}
\Psi^{2}_{2} = {\textstyle \frac{1}{24}}
               \Psi^{0}_{0} \bar{\Psi}{}^{0}_{0}
             + {\textstyle \frac{1}{20}}
               \thopbarnaught (\Psi^{0}_{0}
               \bar{\Psi}{}^{0}_{1} + 4\Psi^{2}_{1})
{\,},
\label{Psitwotwoidentity}
\end{equation}
which, upon substitution into Eq.~(\ref{muthree}),
establishes the chosen form of the coefficient $\mu^{3}$
found in the $\mu$ expansion (\ref{vacexpns}f).

\subsection{Intrinsic geometry of $S(r)$}

In this final appendix subsection we consider the
intrinsic geometry of $S(r)$ and collect radial expansions
along $N_{p}$ for the ``eth'' operator $\thop$, the surface
area element ${\rm d}S$, and the intrinsic curvature
scalar ${\cal R}$ [the Gaussian curvature of $S(r)$
being twice ${\cal R}$]. Again, we consider the
non-vacuum and vacuum cases separately. We obtain the
desired intrinsic-geometry expansions as follows. First,
writing the complex space leg of the null tetrad as\cite{KTW}
\begin{equation}
\delta \equiv
    m^{\mu}\, \partial\!/\!\partial x^{\mu}
    = \xi\, \partial\!/\!\partial\zeta
    + \bar{\eta}\, \partial\!/\!\partial\bar{\zeta}
{\,} ,
\end{equation}
we determine the $r$ expansions along $N_{p}$ for the
tetrad coefficients $\xi$ and $\eta$ from the
{\sc np} commutator equations,\cite{NewmanTod,KTW}
\begin{eqnarray}
D \xi & = & \rho \xi + \sigma \eta
\eqnum{\ref{commutator}a} \\
D\eta & = & \rho\eta + \bar{\sigma} \xi
\eqnum{\ref{commutator}b} \\
\bar{\delta} \xi - \delta \eta & = &
            (\alpha - \bar{\beta}) \xi
          - (\bar{\alpha}  - \beta) \eta {\,} ,
\label{commutator} \eqnum{\ref{commutator}c}
\addtocounter{equation}{1}
\end{eqnarray}
and the radial expansions for the spin coefficients
[Eqs.~(\ref{nonvacexpns}a,b,c,d)
or Eqs.~(\ref{vacexpns}a,b,c,d)
as the case may be]. The given forms of these equations are
particular to how our null frame has been adapted to the
lightcone $N_{p}$.

Let us first consider the $\thop$ operator. With
both Eqs.~(\ref{nonvacexpns}c,d) and the non-vacuum
radial expansions for $\xi$ and $\eta$, we compute the
non-vacuum radial expansion for $\thop$,
\begin{eqnarray}
\thop &\equiv &
             \delta + s (\bar{\alpha} - \beta) \nonumber \\
        & = &
             r^{-1}\thopnaught
           + {\textstyle \frac{1}{6}}r
             \left[\Psi^{0}_{0} \thopbarnaught
           - 4 s \Psi^{0}_{1}
           + \Phi^{0}_{00} \thopnaught
           + 2 s \bar{\Phi}{}^{0}_{10}\right]
           + {\rm O}(r^{2})
             {\,} .
\label{nonvacuumthop}
\end{eqnarray}
Here, as above, $\delta$ is the standard {\sc np} notation
for the directional derivative along $m^{\mu}$,
$s = {\sc sw}(\varphi)$ denotes the spin weight of
some spin-weighted scalar $\varphi$ on which $\thop$ acts,
and in terms of $P = 1 + \zeta \bar{\zeta}$ we define
\begin{equation}
\thopnaught \equiv
\sqrt{{\textstyle \frac{1}{2}}} P \partial/\partial \zeta
+ 2s\bar{\alpha}{}^{0} {\,}.
\label{ethnaught}
\end{equation}
A similar calculation shows that the vacuum $\thop$ operator is
\begin{eqnarray}
       \thop
& \equiv &
       \delta + s (\bar{\alpha} - \beta)
\nonumber \\
&   =    &
       r^{-1} \thopnaught + {\textstyle \frac{1}{6}}r
       \left(\Psi^{0}_{0} \thopbarnaught
     - 4 s \Psi^{0}_{1}\right)
     + {\textstyle \frac{1}{12}}r^2
       \left( \Psi^{1}_{0} \thopbarnaught
     - 5 s \Psi^{1}_{1}\right) \nonumber \\
&        &
     + {\textstyle \frac{1}{360}} r^{3}
       \left[ 7 \Psi^{0}_{0}\bar{\Psi}^{0}_{0}
       \thopnaught + 18 \Psi^{2}_{0} \thopbarnaught
     - 18 s \left(\thopbarnaught \Psi^{2}_{0}\right)
     - 12 s \Psi^{0}_{0} \bar{\Psi}^{0}_{1}\right]
     + {\rm O}(r^{4}) {\,} ,
\label{vacuumthop}
\end{eqnarray}
with $\delta$, $s$, and $\thopnaught$ as before.

Next, we substitute the radial expansions we find for
$\xi$ and $\eta$ into area element of $S(r)$,
\begin{equation}
{\rm d}S =  {\textstyle \frac{1}{2}}{\rm d}\Omega P^{2}
            \left(\frac{\xi\bar{\xi} + \eta \bar{\eta}}{
            \xi\bar{\xi} - \eta \bar{\eta}}\right)
            {\,}.
\label{surfaceelement}
\end{equation}
Here ${\rm d}\Omega$ is the standard unit-radius round sphere
area element. For the non-vacuum scenario
\begin{equation}
      {\rm d}S = {\rm d}\Omega r^{2}
      \left[1 - {\textstyle \frac{1}{3}}r^{2}
      \Phi^{0}_{00} + {\rm O}(r^{3})\right]
{\,} ,
\label{nonvacuumdS}
\end{equation}
while for the vacuum scenario
\begin{equation}
{\rm d}S =
{\rm d}\Omega r^{2}
\left[1 - {\textstyle \frac{1}{90}} r^{4}\Psi^{0}_{0}
\bar{\Psi}^{0}_{0} + {\rm O}(r^{5})\right] {\,} .
\label{vacuumdS}
\end{equation}
These expansions may be checked with the well-known formula
$D({\rm d}S) = -2 \rho {\rm d}S$ for the change in the
surface-area element.\cite{HorowitzSchmidt}

Let us turn to the Gauss curvature of $S(r)$. An order-by-order
examination of the identity
\begin{equation}
       (\bar{\thop}\thop
      - \thop \bar{\thop})
        \varphi
      = {\textstyle \frac{1}{2}}
        {\cal R} \varphi {\,}
\label{Gausscommutator}
\end{equation}
[where {\sc sw}$(\varphi) = 1$, but $\varphi$ is
otherwise arbitrary, and we assume the expansion
$\varphi = \varphi^{0} + r \varphi^{1}
+ r^{2} \varphi^{2} + \cdots$] determines the following
expansion for the intrinsic
curvature of $S(r)$:
\begin{equation}
{\cal R} = 2 r^{-2} + {\cal R}^{0}
         + r {\cal R}^{1} + r^{2} {\cal R}^{2}
         + {\rm O}(r^{3}) {\,}.
\label{Gausscurvature}
\end{equation}
In examining Eq.~(\ref{Gausscommutator}), we must use the
appropriate expansion for $\thop$, Eq.~(\ref{nonvacuumthop})
or Eq.~(\ref{vacuumthop}) as the case may be. For the
non-vacuum scenario, we find the first relevant coefficient
to be
\begin{equation}
      {\cal R}^{0}
    = {\textstyle \frac{2}{3}}\left[\Phi^{0}_{00}
    + \thopnaught \Phi^{0}_{10}
    + \thopbarnaught \bar{\Phi}{}^{0}_{10}
    - 2\thopnaught \bar{\Psi}{}^{0}_{1}
    - 2\thopbarnaught \Psi^{0}_{1}\right] {\,} .
\label{nonvaccurv_0}
\end{equation}
For the vacuum case we need more coefficients. Tedious
but straightforward calculation yields the
set:
\begin{eqnarray}
       {\cal R}^{0}
& = &
     - {\textstyle \frac{4}{3}}
       \left(\thopnaught \bar{\Psi}^{0}_{1}
     + \thopbarnaught \Psi^{0}_{1}\right)
\label{vaccurv} \eqnum{\ref{vaccurv}a}\\
       {\cal R}^{1}
& = &
     - {\textstyle \frac{5}{6}}
       \left( \thopnaught \bar{\Psi}^{1}_{1}
     + \thopbarnaught \Psi^{1}_{1}\right)
\eqnum{\ref{vaccurv}b}\\
       {\cal R}^{2}
& = &
       {\textstyle \frac{1}{45}}
       \Psi^{0}_{0} \bar{\Psi}^{0}_{0}
     - {\textstyle \frac{3}{5}}
       \thopnaught \bar{\Psi}^{2}_{1}
     - {\textstyle \frac{3}{5}}
       \thopbarnaught \Psi^{2}_{1}
     - {\textstyle \frac{17}{90}}
       \thopnaught
       \left(\bar{\Psi}^{0}_{0} \Psi^{0}_{1}\right)
     - {\textstyle \frac{17}{90}}
       \thopbarnaught \left(\Psi^{0}_{0}
       \bar{\Psi}^{0}_{1}\right)  {\,} .
       \eqnum{\ref{vaccurv}c}
\addtocounter{equation}{1}
\end{eqnarray}
We remark that, along the way to obtaining these
coefficients of $\cal R$ from Eq.~(\ref{Gausscommutator}),
one finds terms involving derivatives of the coefficients
of $\varphi$. However, one knows that such terms must vanish,
as the right-hand side of Eq.~(\ref{Gausscommutator}) involves
no derivatives of $\varphi$.
Therefore, one may simply discard these
terms. A detailed examination (order by order in $r$) of the
Bianchi identities (non-vacuum or vacuum as the case may be)
shows that such terms indeed vanish.

\subsection{Vacuum Bianchi Identities}
Consider the vacuum Bianchi identities as given in
in the appendix Ref.~\cite{NewmanTod}. For the situation
at hand, we may simplify these equations by writing
them in terms of $\thop$ rather than $\delta$, and we
do so. Then, considering the equations order by order
in $r$, we find the following at the lowest order in
$r$:
\begin{eqnarray}
\thopnaught \Psi^{0}_{0} = 0
\hspace{8.6mm}
& \hspace{1cm} &
\thopbarnaught \Psi^{0}_{0} = 4 \Psi^{0}_{1}
\label{zerobianchi} \eqnum{\ref{zerobianchi}a,b} \\
\thopnaught \Psi^{0}_{1}
= - {\textstyle \frac{1}{2}}\Psi^{0}_{0}
& &
\thopbarnaught \Psi^{0}_{1} = 3 \Psi^{0}_{2}
\eqnum{\ref{zerobianchi}c,d} \\
 \thopnaught \Psi^{0}_{2} = - \Psi^{0}_{1}
\hspace{2.5mm}
& &
\thopbarnaught \Psi^{0}_{2} = 2 \Psi^{0}_{3}
\eqnum{\ref{zerobianchi}e,f} \\
 \thopnaught \Psi^{0}_{3} =
- {\textstyle \frac{3}{2}}\Psi^{0}_{2}
&  &
\thopbarnaught \Psi^{0}_{3} = \Psi^{0}_{4}
\eqnum{\ref{zerobianchi}g,h} \\
\thopnaught \Psi^{0}_{4} = - 2\Psi^{0}_{3}
& &
\thopbarnaught \Psi^{0}_{4} = 0 {\,} .
\eqnum{\ref{zerobianchi}i,j}
\addtocounter{equation}{1}
\end{eqnarray}
from these one can derive results such as
$\thopnaught\thopbarnaught
\Psi^{0}_{2} = - 3\Psi^{0}_{2}$.


\begin{figure}
\epsfxsize=3.8in
\centerline{\epsfbox{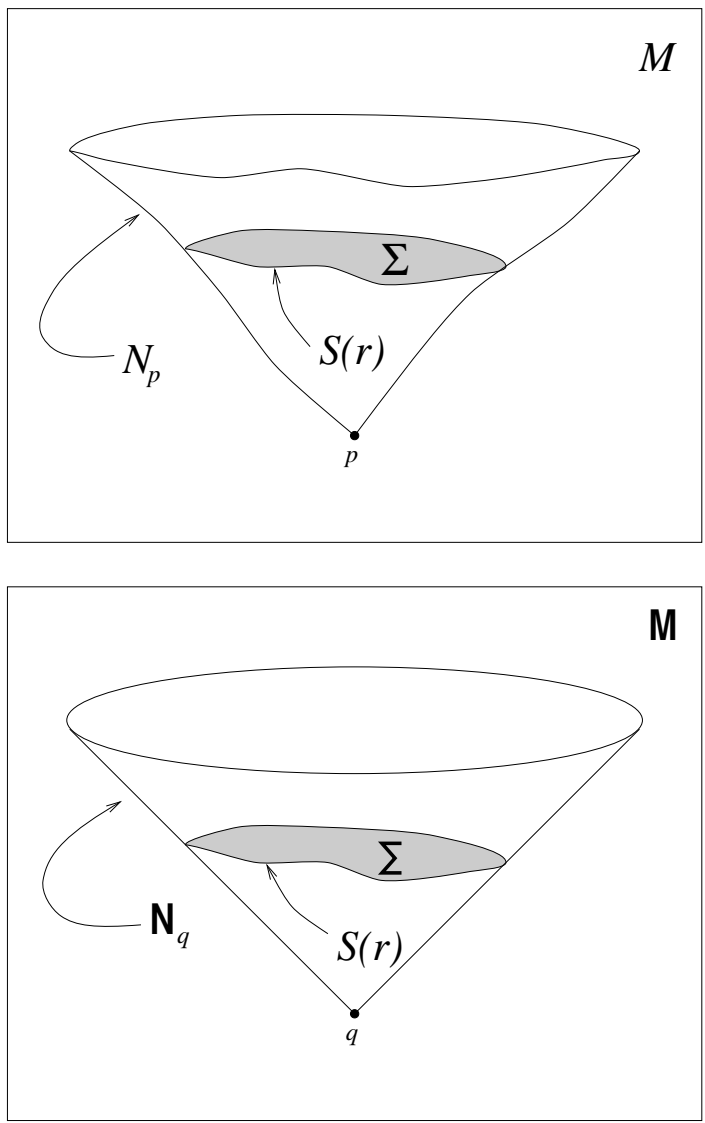}}
\caption{Geometry of the Lightcone Reference.}
{\small Respectively, the top and bottom boxes depict
the physical and reference spacetimes. In the top box
the shaded surface spanning $S(r)$ is the 3-surface
$\Sigma$, while in the bottom box the shaded surface
spanning $S(r)$ is the 3-surface $\referenceSigma$.
Whether viewed as the intersection $N_{p} \bigcap
\Sigma$ or the intersection $\referenceN_{q} \bigcap
\referenceSigma$, the 2-surface $S(r)$ has the same
intrinsic 2-metric.
Our limit construction gives us $\Sigma$, but we must
choose $\referenceSigma$; and, moreover, our choice of
$\referenceSigma$ must be determined solely by
the intrinsic 2-metric of $S(r)$. Our choice
and its physical motivation are described in Subsection 2.A.
However, we note here that, whenever $S(r)$ is at
all distorted from perfect roundness (as it generally will
be), $\referenceSigma$ is not flat Euclidean 3-space
$E^{3}$ (because the intersection of $E^{3}$ with the
genuine lightcone $\referenceN_{q}$ would be a round
sphere). Choosing such a lightcone reference, we
{\em assign} the zero value of the energy to that
(shaded) portion of $\referenceSigma$
contained within $S(r)$, and compute the energy of
(the shaded portion of) $\Sigma$ relative to this
zero-point.}
\end{figure}

\end{document}